\documentclass[traditabstract]{aa} 
\usepackage{txfonts}
\usepackage{epsf}
\usepackage{graphicx}
\usepackage{natbib}
\bibpunct{(}{)}{;}{a}{}{,}

\newcommand{\msol}{\ensuremath{\rm{M}_{\odot}}}
\newcommand{\rsol}{\ensuremath{\rm{R}_{\odot}}}

\newcommand{\teff}{\ensuremath{T_{\rm{eff}}}}
\newcommand{\logg}{\ensuremath{\log g}}
\newcommand{\logy}{\ensuremath{\log y}}
\newcommand{\lheh}{\ensuremath{\log \left(N_{\mathrm{He}}/N_{\mathrm{H}}\right)}}

\newcommand{\twom}{{\scriptsize 2MASS}}
\newcommand{\galex}{{\sc galex}}

\newcommand{\uHz}{\ensuremath{\mu{\rm{Hz}}}}

\newcommand{\ucam}{{Ultracam}}
\newcommand{\busca}{{\sc busca}}
\newcommand{\alfosc}{{\sc alfosc}}
\newcommand{\twin}{{\sc twin}}
\newcommand{\isis}{{\sc isis}}
\newcommand{\kep}{{\em Kepler}}

\newcommand{\mfuss}{{\scriptsize MUCHFUSS}}
\newcommand{\tlusty}{{\sc Tlusty}}
\newcommand{\xtgrid}{{\sc XTgrid}}

\newcommand{\target}{FBS\,0117+396}
\newcommand{\ltarget}{SDSS J012022.94+395059.4}
\defcitealias{geier11a}{{Paper~\sc{i}}}
\defcitealias{geier11b}{{Paper~\sc{ii}}}
\defcitealias{geier11c}{{Paper~\sc{iii}}}

\begin{document}

\title{Binaries discovered by the MUCHFUSS project}
\subtitle{FBS\,0117+396: An sdB+dM binary with a pulsating primary}

\author{
   R.~H.~{\O}stensen \inst{1} \and
   S.~Geier \inst{2,3} \and         
   V.~Schaffenroth \inst{3,4} \and  
   J.~H.~Telting\inst{5} \and       
   S.~Bloemen \inst{1,6} \and         
   P.~N\'emeth \inst{1} \and          
   P.~G.~Beck \inst{1} \and \\      
   R.~Lombaert \inst{1} \and        
   P.~I.~P\'apics \inst{1} \and       
   A.~Tillich \inst{3} \and         
   E.~Ziegerer \inst{3} \and        
   L.~Fox Machado\inst{7} \and      
   S.~Littlefair\inst{8} \and       
   V.~Dhillon\inst{8} \and   \\     
   C.~Aerts \inst{1,6} \and           
   U.~Heber \inst{3} \and           
   P.~F.~L.~Maxted\inst{9} \and     
   B.~T.~G\"ansicke\inst{10} \and    
   T.~R.~Marsh\inst{10} 
}

\institute{
Instituut voor Sterrenkunde, K.\,U.\,Leuven, Celestijnenlaan 200D, 3001 Leuven, Belgium\\
   \email{roy@ster.kuleuven.be}
\and European Southern Observatory, Karl-Schwarzschild-Str.~2, 85748 Garching, Germany
\and Dr.~Karl-Remeis-Observatory \& ECAP, Astronomical Institute, F.-A.\,U.\,Erlangen-N\"urnberg,
     96049 Bamberg, Germany
\and Institute for Astro- and Particle Physics, 
      University of Innsbruck, Technikerstr.\,25/8, 6020 Innsbruck, Austria
\and Nordic Optical Telescope, Rambla Jos\'e Ana Fern\'andez P\'erez 7, 38711 Bre\~na Baja, Spain
\and Department of Astrophysics, IMAPP, Radboud University Nijmegen, 6500 GL Nijmegen, The Netherlands
\and Observatorio Astron\'omico Nacional, 
     Universidad Nacional Aut\'onoma de M\'exico, 
     Ensenada, BC 22860, Mexico
\and Dept of Physics and Astronomy, University of Sheffield,
     Sheffield S3 7RH, UK
\and Astrophysics Group, Keele University, Staffordshire, ST5 5BG, UK
\and Department of Physics, University of Warwick, Coventry CV4 7AL, UK
}

\date{Received 05/06/2013 / accepted 30/09/2013 }
\abstract
{The project Massive Unseen Companions to Hot Faint Underluminous Stars from SDSS ({\sc muchfuss})
 aims to discover subdwarf-B stars with massive compact companions such as overmassive white
dwarfs (M\,$>$\,1.0\,\msol), neutron stars or black holes.
From the 127 subdwarfs with substantial radial-velocity variations discovered in the initial
 survey, a number of interesting objects have been selected for extensive follow-up.

After an initial photometry run with \busca\ revealed that \target\ is photometrically variable both
 on long and short timescales, we chose it as an auxiliary target during a 6-night
 multi-color photometry run with \ucam.
 Spectroscopy was obtained at a number of observatories in order to determine the binary
 period and obtain a radial-velocity amplitude.
After establishing an orbital period of P\,=\,0.252\,d, and removing the signal
associated with the irradiated hemisphere of the M-dwarf companion, we were able to detect
 ten pulsation periods in the Fourier spectrum of the light curve.
 Two pulsation modes are found to have short periods of 337 and 379\,s,
 and at least eight modes are found with periods between 45\,minutes and 2.5\,hours.
This establishes that \target\ is an sdB+dM reflection binary, in which the
primary is a hybrid pulsator, and the first one found with this
particular m\'elange of flavours.
}

\titlerunning{An sdB+dM binary with a pulsating primary}
\authorrunning{\O stensen, Geier, Schaffenroth et al.}

\maketitle

\section{Introduction}

The hot B-type subdwarf (sdB) stars are generally recognised as an extension
of the horizontal branch population where the envelope mass has become too low to
provide significant hydrogen burning. Binary interaction close to the tip of the red-giant
branch is thought to be responsible for the vast majority of the sdB population,
but a variety of end configurations are possible depending on the
initial separation, mass and evolutionary stage of the companion
\citep[see][for a detailed review]{heber09}.

Radial-velocity surveys of sdB stars initiated by \citet{maxted01} and
recently brought closer to completion by \citet{copperwheat11} find that $\sim$50\%\ of all
sdB stars reside in short-period binary systems ($P_{\mathrm{orb}} < 10$\,d).
To date, more than a hundred such short-period binaries are known, and the vast majority of
these have companions with masses compatible with white dwarf (WD) stars.
When the companion is more massive than the subdwarf, or becomes more massive before
mass transfer is terminated, the orbit will expand substantially.
Such orbits are hard to measure, but the companion can be detected spectroscopically or
from infra-red excess. 
\citet{lisker05} found that about a third
of the sdBs in
their survey sample show spectroscopic signatures of main-sequence (MS) companions, while
\citet{reed04}, using \twom\ photometry, inferred that about half of the sdBs in the field
have MS companions. In a recent development, a number of such binaries have
been reported to have orbital periods extending up to 1300 days
\citep{ostensen11d,deca12,ostensen12a,barlow12,vos12}.
The distribution of orbital periods in both the short and long-period binaries are
essential for establishing constraints on the parameters that govern the mass transfer
of theoretical models, and further systems are still sought in order to
improve the statistics and overcome selection biases.

The \mfuss\ project was designed to explore the high-mass end of the short-period
population by pre-selecting stars from the SDSS survey \citep{SDSS} that appeared
to have unusually high radial velocities.
The survey strategy and sample was introduced in \citet[][Paper {\sc i}]{geier11a},
and seven new short-period systems were presented in \citet[][Paper {\sc ii}]{geier11b}.
A new eclipsing system
in which the sdB is accompanied by a substellar companion ($M$\,=\,0.068\,\msol)
with a very short orbital period ($P$\,=\,0.096\,d) was also announced by
\citet[][Paper {\sc iii}]{geier11c}.
In this paper we describe a new binary showing evidence of pulsations from the
sdB primary superimposed on the strong irradiation effect from the tidally locked M-dwarf
companion.  The sdB+dM binaries constitute a fraction of the known short
period binaries. Of the 89 systems listed in Appendix\,A of \citetalias{geier11a} only
19 are of the sdB+dM type, and all of them have periods of less than 0.5 days (19 of 39).
But the statistics from this sample are clearly biased towards those
with the shortest orbital periods, as both photometric and
radial-velocity variations become harder to detect for low-mass
companions as the period increases.

Subdwarf-B stars have been found to pulsate, first in rapid pressure ($p$-)modes
with periods between $\sim$1 and 10 minutes by \citet{kilkenny97}, and 
later in slower gravity ($g$-)modes with periods between $\sim$1/2 and 2 hours
by \citet{green03}. Such pulsators are commonly referred to as sdB pulsators (sdBVs),
and several flavours are recognised. Conventionally, we designate pulsating stars
according to the variable star names of the prototypical objects,
so that the short-period pulsators are known as V361-Hya stars, the long-period pulsators
are V1093-Her stars, and hybrid stars displaying both types of pulsations simultaneously
are known as DW-Lyn stars.  Of the sdB+dM binaries,
only a handful are known to have pulsating primaries.  The most well known is
NY\,Vir \citep{kilkenny98}, which being an eclipsing system permits one to compare
detailed constraints obtained from analysing orbital spectroscopic and
photometric effects \citep{vuckovic07} with mass and radius estimates obtained
from asteroseismic modelling of the pulsator \citep{charpinet08,vangrootel13}.
Another one of the classic sdBV stars, \object{V1405\,Ori} \citep{koen99b}, was recently
discovered to be the second sdBV+dM binary by \citet{reed10b}, and \object{HE\,0230--4323}
was reported to be a short-period sdB pulsator with a strong reflection effect by
\citet{kilkenny10c}. Up to now these three were
the only examples of short-period sdB pulsators with M-dwarf companions.
And for long-period sdBVs only a single case had been reported; JL\,87
\citep{koen09}, until the recent advent of space-based photometry of sdB stars.
A sample of compact-pulsator candidates was monitored with
the \kep\ spacecraft \citep{ostensen10b,ostensen11b}, and revealed a number
of long-period sdBV stars with M-dwarf companions, all of which have pulsation
amplitudes too low to be easily detected from the ground. The most spectacular one is
\object{2M1938+4603} \citep{2m1938}, which is an eclipsing binary with a period of 0.126\,d,
in which the primary pulsates with an unusually high number of low-amplitude modes
covering both long and intermediate period ranges.
Two non-eclipsing sdB+dM binaries in which the primaries are long-period
pulsators of the V1093-Her type were presented by \citet{kawaler10b}, 
revealing orbital periods of 0.443 and 0.395 days, a range which is difficult to
discover with ground based observations.
Recently, a fourth sdB+dM binary with a pulsating primary was revealed in \kep\
data by \citet{pablo11}. This system with an orbital period of 0.399\,d is very similar
to the two systems presented by \citet{kawaler10b}, but is a member of the open cluster
\object{NGC 6791}. Interestingly, \citet{pablo11} were able to infer a rotation period
of 9.63\,days of the primary from the characteristic even splitting of the pulsation periods,
thereby demonstrating that the rotation of the primary is not synchronised with the orbital
period. More recently \citet{pablo12} have demonstrated that the same is the case for both
of the long-period pulsators of \citet{kawaler10b}. Long-period pulsations have also been
reported from ground-based studies of the sdB+dM binary \galex\,J0321+4727, which has an
orbital period of 0.266\,d \citep{kawka12}.
Since the exceptionally rich pulsation spectrum of 2M1938+4603 can only be explained if the
sdB primary is rotating with a period close to the orbital, it is clear that tidal synchronisation
of the primary in sdB+dM binaries is efficient at periods of 1/8\,d but not at 2/5\,d.
The study of pulsators in systems with intermediate periods, say around 1/4 to 1/3\,d is therefore
of interest in order to determine at which point the tidal influence becomes sufficient to
enforce synchronisation.
The system presented here, \target, has an orbital period of 1/4\,d, and may therefore be
significant in this respect.

\begin{figure}
\includegraphics[width=\hsize]{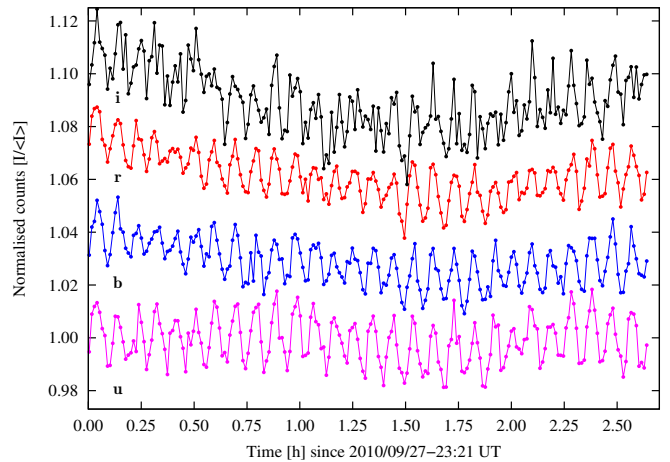}
\caption{
Discovery light curve of \target\ obtained with \busca.
The $b$, $r$ and $i$ curves have been shifted upward by 0.03$n$ for
clarity.
}
\label{fig:busca_lc}
\end{figure}

\section{Discovery data}


\target\ occurs in the Sloan Digital Sky Survey as \ltarget\ with {\em ugriz} magnitudes
15.051(4), 15.158(3), 15.638(4), 15.977(4), 16.265(7).
It was originally designated \object{FBS\,B12}
in the First Byurakan Sky Survey \citep{abrahamian90} and classified as a hot
subdwarf (sdB--O). It is close to the confusion limit of \twom, and it is listed
with $J$ and $K$ magnitudes of 16.02(8), and 15.9(2) and only an upper limit on $K_s$
\citep{twomass}.
\target\ is also found in the \galex\ survey, and the magnitudes from the medium imaging
survey are FUV, NUV = 14.422(6), 14.586(4).
The extinction towards infinity in the direction of
\target\ ($l$,$b$\,=\,128.9494, --22.6839) according to the dust maps
of \citet{schlegel98} is $E(B-V)$\,=\,0.053.

\begin{figure}
\includegraphics[width=\hsize]{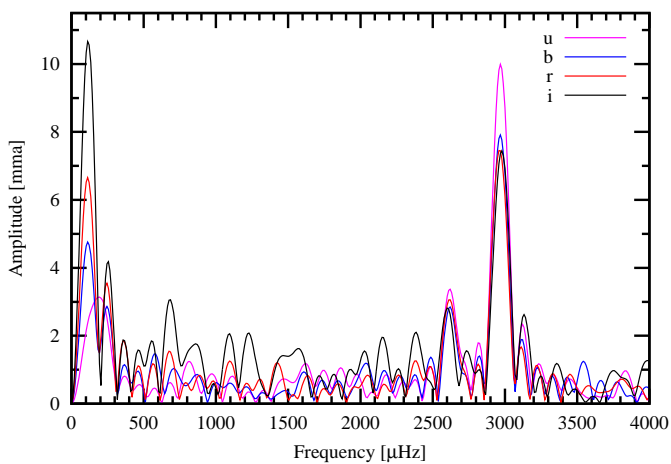}
\caption{
Amplitude spectrum of the discovery light curves obtained with \busca,
as shown in Fig.~\ref{fig:busca_lc}.
}
\label{fig:busca_ft}
\end{figure}

\begin{table*}[t!]
\caption[]{Log of observations.}
\label{tbl:obslog} \centering
\begin{tabular}{llllll}
\hline\hline \noalign{\smallskip}
Year/Month/Days & Telescope & Instrument & Passbands & Resolution & P.I., Observer \\
\noalign{\smallskip} \hline \noalign{\smallskip}
2004/12/12 & SDSS & Spectro & 3800--9200 & $\sim$2000 & Archive data \\
2005/12/08 & SDSS & Spectro & 3800--9200 & $\sim$2000 & Archive data \\
2009/08/27--28 & WHT 4.2m & \isis & 3700--5000 & $\sim$3400 & SG, AT \\
2010/08/28--31 & WHT 4.2m & \isis & 3700--5000 & $\sim$2400 & CA, RH\O \\
2011/09/17--22 & {\scriptsize NOT} 2.5m & \alfosc & 3500--5100 & $\sim$2200 & JHT, Service \\
2011/11/15--17 & {\scriptsize CAHA} 3.5m & \twin & 3700--5400 & $\sim$4000 & SG \\
2011/11/20--21 & {\scriptsize NOT} 2.5m & \alfosc & 3500--5100 & $\sim$2200 & JHT, Service \\
\noalign{\smallskip} \hline \noalign{\smallskip}
2010/09/27     & {\scriptsize CAHA} 2.2m & \busca & {\em u,b,r,i} & & SG, VS \&\ EZ \\
2011/08/13--14 & Mercator 1.2m & Merope {\sc ii} & White light & & RH\O , PP \\
2011/08/22     & Mercator 1.2m & Merope {\sc ii} & White light & & RH\O , RL \\
2011/08/28--09/03 & Mercator 1.2m & Merope {\sc ii} & White light & & RH\O , PB \\
2011/08/16--21 & WHT 4.2m & \ucam & {\em u,g,r} & & CA, RH\O\ \&\ SL \\
2011/08/27--29 & OAN-SPM 0.84m & Mexman & White light & & LFM \\
\noalign{\smallskip} \hline
\end{tabular}
\end{table*}
The SDSS spectroscopic archive contains observations of this target at two epochs, the
first on December 12, 2004 (MJD\,=\,53351) and again one year later.
We downloaded the reduced SDSS 2D bundles of spectra for the two plates, and found that
the first one (plate 2043) contains three individual 900\,s integrations, and the second
(plate 2336) contains five integrations with exposure times varying from 900 to 1500\,s,
indicating that the last plate was taken in rather poor conditions. Still, since our
target is one of the brightest on the plate, the S/N is excellent ($>$50) in all the
individual integrations. From these eight spectra we found radial velocities that varied
from close to zero to more than 100 km/s between the two epochs, and a trend of more than
30 km/s in the second epoch run, clear evidence that \target\ is a short-period
binary. It was therefore flagged as a priority target for follow-up \citepalias[see][]{geier11a},
and observed with the {\sc isis} spectrograph on the William Herschel Telescope on
La Palma, Canary Islands, in August 2009.
We observed it again with WHT/{\sc isis} a year later to further constrain the orbit solution. 

In September 2010 a 2.5\,h run of \target\ was obtained
with the four-channel CCD camera \busca\ \citep{busca},
mounted on the {\scriptsize CAHA} 2.2-m telescope.
The multi-colour light curve is shown in 
Fig.\,\ref{fig:busca_lc}, 
and the corresponding Fourier transform in Fig.~\ref{fig:busca_ft}.
Note that \busca\ observations are done without any filters, but with
passbands defined by the dichroics which splits the input
beam roughly at 4400, 5400 and 7300\,\AA, providing four channels designated
as {\em u,b,r,i} in Fig.~\ref{fig:busca_lc}.
It is immediately clear from the photometry that
the target is a rapid pulsator with a pulsation semiamplitude of about
1\%\ at a period of $\sim$340\,s, as reported briefly by \citet{geier12a}.
The pulsations appear to have a higher
amplitude at shorter wavelengths, as one would expect for low-order
pulsation modes in sdBVs \citep{asteroseismology}.
One can also notice that there is a
long-term trend with a minimum during the run, which is most
significant in the $i$-band, and practically absent in $u$. This
is consistent with an irradiation effect in a close sdB+dM binary.

In order to confirm these suspicions and get a firm orbital period and
velocity amplitude for the binary, we targeted \target\ at several
opportunities in 2011. Service-mode observations netting 17 spectra 
were obtained on two occasions with \alfosc\ on the
2.5-m Nordic Optical Telescope ({\scriptsize NOT}), also on La Palma,
and another 17 spectra were obtained with the \twin\ spectrograph
on the 3.5-m telescope at the German-Spanish Astronomical Center at Calar Alto
({\scriptsize CAHA}) in Spain. Details for these observations
are given in the upper half of Table~\ref{tbl:obslog}.

During a photometric campaign on the peculiar low-gravity sdB pulsator
\object{\galex\,J201337+092801} \citep{ostensen11a} in August 2011, we
took the opportunity to observe \target\ as a secondary target when the
priority target started to get low in the sky. The photometric observations 
are listed in the lower half of Table~\ref{tbl:obslog}, and will be
analysed in detail in Section\,\ref{sect:phot}.

\begin{table*}[t]
\caption[]{Radial velocity data.}
\label{tbl:rvs}\centering\small
\begin{tabular}{crrrrr|crrrrr}
\hline\hline \noalign{\smallskip}
 HJD & RV$_{\mathrm{obs}}$ & RV$_{\mathrm{hel}}$ & RV$_{\mathrm{K}}$ 
 & $\sigma_{\mathrm{RV}}$ & $\sigma_{\mathrm{K}}$ & 
 HJD & RV$_{\mathrm{obs}}$ & RV$_{\mathrm{hel}}$ & RV$_{\mathrm{K}}$ 
 & $\sigma_{\mathrm{RV}}$ & $\sigma_{\mathrm{K}}$ \\
 --2450000 & [km/s] & [km/s] & [km/s] & [km/s] & [km/s] &
 --2450000 & [km/s] & [km/s] & [km/s] & [km/s] & [km/s] \\
\noalign{\smallskip} \hline \noalign{\smallskip}
3351.57631 & 5.01 & -14.74 & -14.74 & 19.18 & 8.66 & 
5439.65877 & -15.02 & 6.50 & -87.92 & 11.80 & 2.00 \\ 
3351.58874 & 11.71 & -8.07 & -8.07 & 18.92 & 8.66 & 
5439.66930 & -14.47 & 7.02 & -66.79 & 10.34 & 2.00 \\ 
3351.60131 & 3.82 & -15.99 & -15.99 & 19.85 & 8.66 & 
5439.67983 & -23.42 & -1.95 & -55.16 & 11.18 & 2.00 \\ 
3712.57476 & -86.28 & -104.58 & -75.32 & 20.76 & 6.70 & 
5439.69036 & -35.20 & -13.75 & -47.12 & 7.43 & 2.00 \\ 
3712.59380 & -85.61 & -103.95 & -74.69 & 20.60 & 6.70 & 
5822.45519 & -101.27 & -84.69 & -82.70 & 7.77 & 5.50 \\ 
3712.61486 & -68.61 & -86.99 & -57.73 & 20.50 & 6.70 & 
5822.48965 & -64.93 & -48.40 & -46.36 & 6.62 & 5.50 \\ 
3712.63068 & -61.57 & -79.99 & -50.73 & 21.04 & 6.70 & 
5822.51908 & -34.36 & -17.89 & -15.79 & 6.91 & 5.50 \\ 
3712.64250 & -51.59 & -70.04 & -40.78 & 21.19 & 6.70 & 
5822.53630 & -41.10 & -24.66 & -22.53 & 7.39 & 5.50 \\ 
5070.66930 & -82.62 & -60.27 & -83.11 & 10.67 & 1.63 & 
5826.56409 & -18.80 & -3.89 & 1.29 & 8.27 & 5.50 \\ 
5070.67651 & -86.93 & -64.59 & -80.56 & 10.70 & 1.63 & 
5826.60696 & -24.50 & -9.69 & -4.41 & 13.31 & 5.50 \\ 
5070.68372 & -86.34 & -64.02 & -73.11 & 11.98 & 1.63 & 
5826.65140 & -62.10 & -47.40 & -42.01 & 9.59 & 5.50 \\ 
5070.69424 & -93.34 & -71.04 & -70.96 & 8.45 & 1.63 & 
5881.38563 & -65.27 & -74.54 & -95.78 & 15.03 & 3.05 \\ 
5070.70063 & -94.55 & -72.27 & -65.31 & 9.24 & 1.63 & 
5881.39530 & -77.99 & -87.29 & -108.50 & 16.16 & 3.05 \\ 
5070.70784 & -101.39 & -79.12 & -65.29 & 6.18 & 1.63 & 
5881.40882 & -55.42 & -64.73 & -85.93 & 12.81 & 3.05 \\ 
5071.53483 & -12.35 & 10.04 & -24.27 & 7.65 & 1.63 & 
5881.41086 & -45.95 & -55.28 & -76.46 & 19.12 & 3.05 \\ 
5071.54204 & -10.57 & 11.81 & -16.39 & 8.63 & 1.63 & 
5881.42160 & -62.24 & -71.60 & -92.75 & 16.95 & 3.05 \\ 
5071.54925 & -9.84 & 12.52 & -9.56 & 10.38 & 1.63 & 
5881.43146 & -59.12 & -68.50 & -89.63 & 12.58 & 3.05 \\ 
5071.56058 & -15.73 & 6.62 & -6.31 & 11.47 & 1.63 & 
5881.43953 & -40.86 & -50.26 & -71.37 & 18.59 & 3.05 \\ 
5071.56779 & -24.13 & -1.80 & -8.61 & 8.52 & 1.63 & 
5881.44752 & -40.61 & -50.02 & -71.12 & 11.91 & 3.05 \\ 
5071.57500 & -35.19 & -12.87 & -13.57 & 7.52 & 1.63 & 
5881.45686 & -23.46 & -32.90 & -53.97 & 17.45 & 3.05 \\ 
5071.63409 & -61.12 & -38.91 & -63.13 & 6.84 & 4.00 & 
5881.46488 & -12.73 & -22.18 & -43.24 & 15.40 & 3.05 \\ 
5071.64477 & -81.25 & -59.07 & -80.21 & 8.09 & 4.00 & 
5881.56221 & -1.18 & -10.82 & -31.69 & 16.12 & 3.05 \\ 
5436.62102 & -66.71 & -44.42 & -73.29 & 9.41 & 1.41 & 
5881.57240 & -18.53 & -28.18 & -49.04 & 12.41 & 3.05 \\ 
5436.63085 & -65.85 & -43.58 & -72.43 & 7.60 & 1.41 & 
5882.52260 & 5.31 & -4.71 & -25.20 & 10.98 & 3.05 \\ 
5436.64138 & -55.27 & -33.02 & -61.85 & 9.38 & 1.41 & 
5882.54198 & 17.34 & 7.29 & -13.17 & 8.85 & 3.05 \\ 
5436.65191 & -43.17 & -20.94 & -49.75 & 8.78 & 1.41 & 
5882.55276 & 41.87 & 31.80 & 11.36 & 15.48 & 3.05 \\ 
5436.66244 & -34.20 & -11.99 & -40.78 & 10.70 & 1.41 & 
5882.56747 & -7.66 & -17.75 & -38.17 & 11.11 & 3.05 \\ 
5436.67430 & -26.62 & -4.45 & -33.20 & 10.07 & 1.41 & 
5886.33142 & -13.73 & -25.16 & -11.93 & 7.46 & 3.89 \\ 
5436.68482 & -22.20 & -0.04 & -28.78 & 8.62 & 1.41 & 
5886.39885 & -66.14 & -77.40 & -64.34 & 9.89 & 3.89 \\ 
5436.69535 & -14.05 & 8.08 & -20.63 & 7.07 & 1.41 & 
5886.45854 & -62.83 & -74.24 & -61.03 & 7.53 & 3.89 \\ 
5436.70588 & -5.54 & 16.57 & -19.75 & 6.83 & 4.00 & 
5886.50125 & -30.91 & -42.46 & -29.11 & 10.41 & 3.89 \\ 
5437.69600 & -11.23 & 10.68 & -32.30 & 9.88 & 4.00 & 
5886.54449 & -6.50 & -18.15 & -4.70 & 11.38 & 3.89 \\ 
5438.71102 & -4.38 & 17.26 & -36.88 & 6.51 & 2.31 & 
5886.58586 & -1.74 & -13.49 & 0.06 & 12.86 & 3.89 \\ 
5438.72154 & -4.85 & 16.77 & -20.58 & 6.10 & 2.31 & 
5886.62683 & -47.49 & -59.32 & -45.69 & 13.34 & 3.89 \\ 
5438.73207 & -9.91 & 11.69 & -8.87 & 6.59 & 2.31 & 
5886.67472 & -78.84 & -90.73 & -77.04 & 11.96 & 3.89 \\ 
5439.65877 & -15.02 & 6.50 & -87.92 & 11.80 & 2.00 & 
 & & & & & \\ 
\noalign{\smallskip} \hline
\end{tabular}
\end{table*}

\section{Spectroscopic follow-up observations}\label{sect:rv}

The spectroscopic observations from WHT, {\scriptsize CAHA}\
and {\scriptsize NOT} were extracted
and processed with standard {\scriptsize IRAF}\footnote{
{\sc iraf} is distributed by the National Optical Astronomy Observatory;
see http://iraf.noao.edu/.} tasks. Radial velocities
were computed with {\tt fxcor}, by cross-correlating with
a synthetic template derived from a suitable mean spectrum of the
target. During the processing we encountered a problem with the
calibrations for some of the WHT data. It turned out that for some of
the runs, instead of obtaining an arc at the beginning and at the end of
the sequence, in order to calibrate the wavelength solution as interpolations
of the two arcs, for the sequences obtained in 2009 only a single arc
was taken close to the middle of each set, and for the last sequence
taken in 2010, only a single arc was obtained at the beginning. When
checking the position of the interstellar \ion{Ca}{ii} K-line at 3933\,\AA, it was
found to move by more than 30 km/s over sequences lasting about one hour,
much more than anticipated. Fortunately, our spectra have fairly
high S/N and the K-line is quite strong and can be measured quite
reliably. We therefore computed the position of the K-line for each
spectrum, and for each set of observations we computed either an
average velocity correction from the whole set (for those with
good wavelength solutions), or a linear fit (for those with wavelength
calibration problems). Note that this correction with respect to an
insterstellar line circumvents the need for a heliocentric correction,
but increases the uncertainty of each measurement by the 
error on the K-line measurement.
In Table~\ref{tbl:rvs} the radial velocities are listed for each
spectroscopic observation as measured by the cross-correlation procedure
(RV$_{\mathrm{obs}}$) together with their values as corrected with
the heliocentric velocity (RV$_{\mathrm{hel}}$), and with
respect to the K-line (RV$_{\mathrm{K}}$) when taken to be constant
at its mean value of 3933.064\,\AA.
The errors listed as $\sigma_{\mathrm{RV}}$ and $\sigma_{\mathrm{K}}$ in
Table\,\ref{tbl:rvs} are respectively {\sc verr} from the {\sc iraf} {\tt fxcor} task
and the RMS error reported when fitting a Gaussian profile to the K-line with the
Levenberg-Marquardt fitting method built into the {\sc iraf} {\tt splot} task.

\begin{figure}
\includegraphics[width=\hsize]{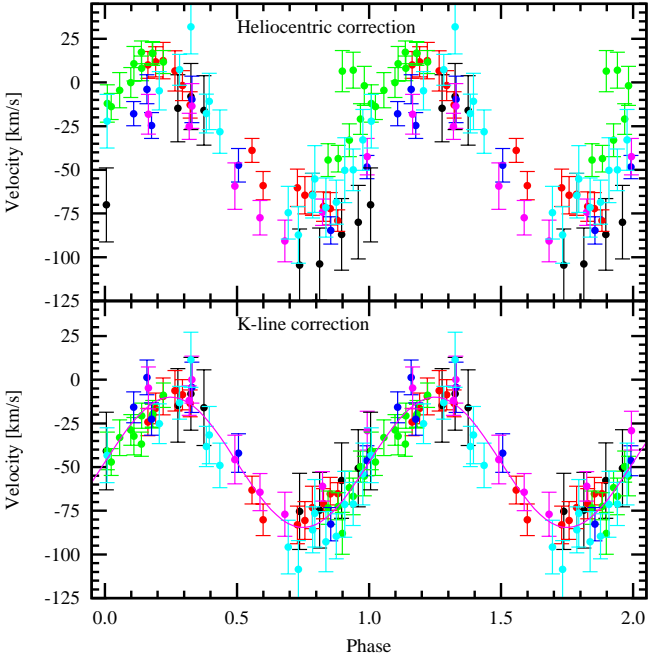}
\caption{
Radial velocities folded on the best fit orbital period.
The upper panel shows the results when using the standard
Heliocentric correction, and the lower panel when using
the \ion{Ca}{ii} K-line to correct the zero points.
The symbols indicate the origin of the data points, {\em i.e.}
SDSS (black), WHT 2009 (red) and 2010 (green),
{\scriptsize NOT} 2011/09 (blue) and 2011/11 (magenta), and {\scriptsize CAHA} (cyan). 
}
\label{fig:rvplot}
\end{figure}


The best fitting solution, assuming a circular orbit,
for the K-line corrected set of velocities is:
\begin{eqnarray*}
T_0   & = & 2455439.6843 \pm 0.0022\,{\mathrm d}, \\
P     & = & 0.252013 \pm 0.000013\,{\mathrm d}, \\
\gamma& = & -47.3 \pm 1.3\,{\mathrm{km/s}}, \\
K_1   & = & 37.3 \pm 2.8\,{\mathrm{km/s}},
\end{eqnarray*}
where $T_0$ is the Heliocentric Julian Date (HJD) of zero phase
(where the primary is at the closest point to the observer),
$P$ is the orbital period, $\gamma$ is the velocity of the binary
system relative to the sun, and $K_1$ is the semi-amplitude of
the radial velocity variation.

The mass function for this system is then
\begin{eqnarray*}
f(m) & = & \frac{(M_2 \sin i)^3}{(M_1+M_2)^2}
    = 0.00135\,\msol,
\end{eqnarray*}
which implies that for a typical $M_1$\,=\,0.47\,\msol,
and a secondary at the substellar limit ($M_2$\,=\,0.075\,\msol), 
the inclination angle must be $i$\,=\,80$^{\circ}$, which is ruled
out by the absence of eclipses. A more massive secondary
would imply a lower angle, {\em i.e.} a mass of 0.1\,\msol\ corresponds
to $i$\,=\,50$^{\circ}$, which is permitted.

\begin{table}[h!]
\caption[]{Physical parameters derived from the detrended mean spectra.}
\label{tbl:physpar}
\centering
\begin{tabular}{lccc} \hline\hline \noalign{\smallskip}
Spectrum   & \teff & \logg & \logy \\
           & [kK]  & [dex] & [dex] \\ 
\noalign{\smallskip} \hline \noalign{\smallskip}
SDSS       & 27.8\,$\pm$\,0.3 & 5.32\,$\pm$\,0.03 & --3.01\,$\pm$\,0.07 \\
WHT1       & 27.9\,$\pm$\,0.1 & 5.45\,$\pm$\,0.01 & --3.03\,$\pm$\,0.04 \\
WHT2       & 28.6\,$\pm$\,0.1 & 5.43\,$\pm$\,0.01 & --3.02\,$\pm$\,0.02 \\
NOT1       & 28.5\,$\pm$\,0.2 & 5.42\,$\pm$\,0.03 & --3.01\,$\pm$\,0.06 \\
CAHA       & 28.7\,$\pm$\,0.3 & 5.41\,$\pm$\,0.03 & --3.04\,$\pm$\,0.10 \\
NOT2       & 28.8\,$\pm$\,0.3 & 5.42\,$\pm$\,0.04 & --2.99\,$\pm$\,0.07 \\
\noalign{\smallskip} \hline \noalign{\smallskip}
Adopted  & 28.5\,$\pm$\,0.1 & 5.42\,$\pm$\,0.01 & --3.02\,$\pm$\,0.05 \\
\noalign{\smallskip} \hline
\end{tabular}
\end{table}

\subsection{Physical parameters}\label{sect:spfit}

We constructed mean spectra for each run by shifting the observed
spectra according to the determined orbit.
The hydrogen and helium lines of each such 
detrended mean spectrum were then fitted with a grid of synthetic spectra 
calculated from fully line blanketed LTE model atmospheres assuming 
solar metalicity \citep{heber00}.
The results of these fits are listed in Table~\ref{tbl:physpar}.
Note that formal fitting errors stated in the table do not account for
systematic effects inherent in the models, so we generously increase
the errors when stating
\teff\,=\,28\,500$\pm$500\,K, \logg\,=\,5.42$\pm$0.10,
and \lheh\,=\,--3.05$\pm$0.10.

\begin{table}[b]
\caption[]{Fitted lines with equivalent widths larger then 50\,m\AA.}
\label{tbl:lines}
\centering
\begin{tabular}{lcr|lcr} \hline\hline \noalign{\smallskip}
Ion   & Wavelength & $W_\lambda$ & Ion   & Wavelength  & $W_\lambda$  \\
      & [\AA]      & [m\AA] & & [\AA]      & [m\AA] \\
\noalign{\smallskip} \hline \noalign{\smallskip}
He {\sc i}  & 3888.65  & 148.3 & O {\sc ii}   & 4649.14 & 62.4  \\
He {\sc i}  & 4026.19  & 78.4  & Mg {\sc ii}  & 4481.13 & 53.2 \\
He {\sc i}  & 4471.47  & 125.6 & Si {\sc iii} & 4567.84 & 54.2  \\
He {\sc i}  & 4471.49  & 136.1 & Si {\sc iii} & 4567.84 & 54.2  \\
He {\sc i}  & 4471.68  & 51.4  & Fe {\sc iii} & 4137.76 & 50.6  \\
He {\sc i}  & 4921.93  & 138.1 & Fe {\sc iii} & 4164.73 & 57.2  \\
He {\sc i}  & 5015.68  & 65.4  & Fe {\sc iii} & 4310.36 & 51.2  \\
N {\sc ii}  & 3994.99  & 83.5  & Fe {\sc iii} & 4372.82 & 50.0  \\
N {\sc ii}  & 4241.79  & 57.4  & Fe {\sc iii} & 4419.60 & 55.0  \\
N {\sc ii}  & 4630.54  & 78.8  & Fe {\sc iii} & 5127.37 & 56.8  \\
N {\sc ii}  & 5001.13  & 63.0  &              &         &       \\
N {\sc ii}  & 5001.47  & 60.7  &              &         &       \\
N {\sc ii}  & 5005.15  & 78.5  &              &         &       \\
\noalign{\smallskip} \hline
\end{tabular}\end{table}

\begin{figure}
\includegraphics[width=\hsize]{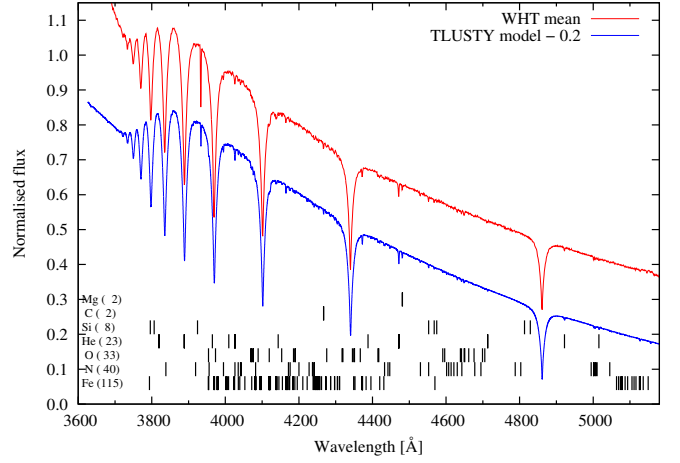}
\caption{
The sum of the 17 WHT spectra from 2010 after correcting for the orbital velocity.
The S/N in this mean spectrum peaks at $\sim$350, and many weak metal lines
can be distinguished in addition to the Balmer lines and \ion{He}{i}
lines at 4472 and 4026\,\AA.
The slope of the observed spectrum was corrected by dividing with an instrument
response function derived from a flux standard to put it on a $F_\lambda$ grid.
The absolute units are arbitrary.
Shifted down by 0.2 is the model fit
computed with \tlusty/\xtgrid. The final parameters for this model fit are given 
in Table\,\ref{tbl:abund}.
}
\label{fig:wht_sp}
\end{figure}

We also made a fit of the mean WHT spectrum from 2010
using the \tlusty\ code \citep{tlusty},
which is a general NLTE model atmosphere code for early-type stars, treating metal
line blanketing by opacity sampling. We applied the
\xtgrid\ fitting program \citep{nemeth12}, and including the
first 30 elements in the fit.
The fitting algorithm is a standard $\chi^2$-minimization technique, where
the procedure starts from a detailed model and with successive approximations
along the steepest gradient of the $\chi^2$ surface, it converges on a solution.
The procedure was designed to fit the entire spectrum and not just selected
lines, so as to account for line blanketing. However the fit is still
primarily driven by the Balmer lines with contributions from the
strongest metal lines (listed in Table~\ref{tbl:lines}).
The best fit was found with \teff\,=\,29370\,K
and \logg\,=\,5.48\,dex, and errors and abundances for the elements
that were found to be significant as listed in Table\,\ref{tbl:abund}.
Parameter errors were determined by changing the model in one dimension\
until the critical $\chi^2$-value associated with the probability level at the
given number of free parameters was reached.
The resulting fit is shown together with the mean spectrum 
in Fig.\,\ref{fig:wht_sp}.
While the fit converged with a high rotational velocity for the sdB,
$v\sin{i}=76\pm10\ \mathrm{km/s}$, this value is most likely severely exaggerated due
to smearing introduced by the merging of many spectra with poor wavelength
calibration, so we do not consider it a real measurement.
Note that the higher temperature of the NLTE model is consistent
with the typical shift of about +1000\,K
when going from LTE to NLTE
\citep[see][and references therein]{heber09}.
The abundances show that iron is solar whereas the other elements are subsolar,
which is a typical pattern in sdB stars \citep[see e.g.][]{geier13a}.

\begin{table}[b]
\caption[]{Parameters for the fit shown in Fig.\,\ref{fig:wht_sp}, with
the solar abundances from \citet{asplund09} provided for comparison.}
\label{tbl:abund}
\centering
\begin{tabular}{llrrrr} \hline\hline \noalign{\smallskip}
Parameter                      & Value   & $+1 \sigma$ & $-1 \sigma$ & Solar & Unit\\
\noalign{\smallskip} \hline \noalign{\smallskip}
\teff\                           & 29370 & 60 & 250 & & K\\
\logg\                           &  5.484 & 0.020 & 0.013 & & dex \\
$\log n(\mathrm{He})/n(\mathrm{H})$\ &--2.99 & 0.11 & 0.13 & --1.07 & dex \\
$\log n(\mathrm{C})/n(\mathrm{H}) $\ &--4.92 & 0.75 & 0.57 & --3.57 & dex \\
$\log n(\mathrm{N})/n(\mathrm{H}) $\ &--4.78 & 0.30 & 0.20 & --4.17 & dex \\
$\log n(\mathrm{O})/n(\mathrm{H}) $\ &--4.61 & 0.26 & 0.30 & --3.31 & dex \\
$\log n(\mathrm{Mg})/n(\mathrm{H})$\ &--5.32 & 0.48 & 0.65 & --4.40 & dex \\
$\log n(\mathrm{Si})/n(\mathrm{H})$\ &--5.84 & 0.51 & 0.69 & --4.49 & dex \\
$\log n(\mathrm{Fe})/n(\mathrm{H})$\ &--4.40 & 0.16 & 0.20 & --4.50 & dex \\
\noalign{\smallskip} \hline
\end{tabular}
\end{table}

\section{High-speed photometric observations}\label{sect:phot}

The pulsations in \target\ were revealed during a short exploratory run
with \busca. The amplitude spectra corresponding to the light curve in
Fig.\,\ref{fig:busca_lc}, is plotted in Fig.\,\ref{fig:busca_ft}.
The most significant peak is found at 2970\,\uHz, with a second lower peak
detected at 2610\,\uHz. A possible third peak may be present at 3120\,\uHz.

\begin{figure*}
\centering
\includegraphics[width=15cm]{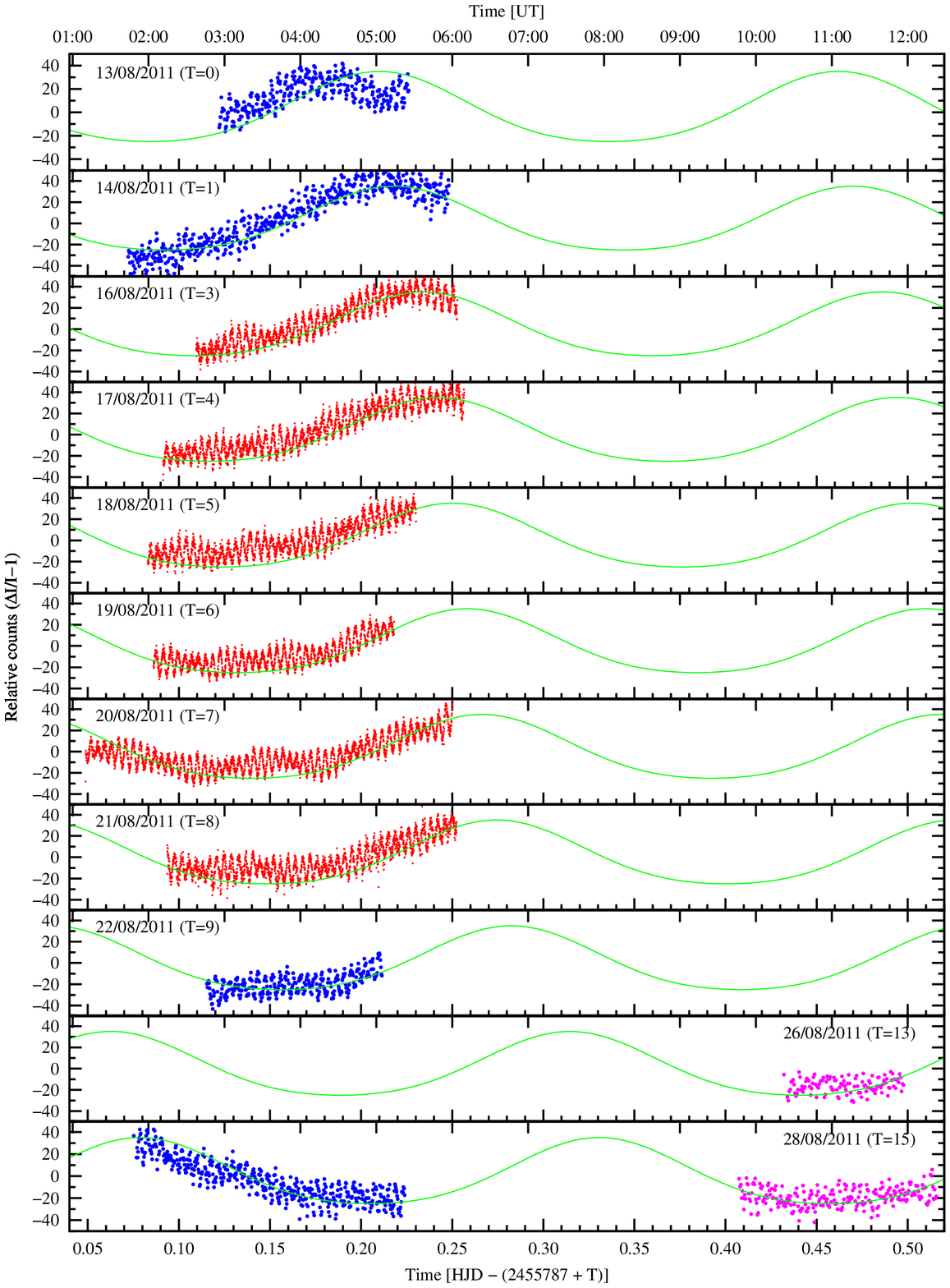}
\caption{
White-light data for the whole campaign. Red points mark \ucam,
blue points Merope\,{\sc ii}, and magenta points Mexman data.
The solid curve corresponds to the simple model given by Eq.~\ref{eq:lc},
with the period and phase of the radial-velocity solution.
}
\label{fig:lc_all}
\end{figure*}

During the photometric campaign of 2011 we obtained about four hours of
data on \target\ on each of six consecutive nights with \ucam\ \citep{ultracam}.
Additionally, we obtained supporting
data from two other observatories on five adjacent nights, four runs with the
Merope\,{\sc ii} imager on the 1.2-m Mercator telescope \citep{ostensen10},
and three runs with Mexman on the 0.84-m telescope
at San Pedro M\'artir (SPM) observatory in Mexico, of which one night was too cloudy
to be useful (see log of observations in Table\,\ref{tbl:obslog}).
The reduced photometric data is shown in Fig.\,\ref{fig:lc_all}.
Observing conditions were good on these nights, with clear sky but variable seeing
on some nights.
While \ucam\ was observing in three bands simultaneously, the smaller supporting
telescopes used white-light (unfiltered) observations. White-light data from
\ucam\ was generated by summing the $r$+$g$+$u$ photometry.
The short-period pulsations are obvious, and the long-period pulsations can also be
discerned as excursions with a timescale of about 0.05d. The orbital frequency
is also quite clear, at least when the model plotted with a continuous curve
is used as a guide. Since the orbital frequency is close to four cycles per day,
it only shifts by 11.6 minutes every night.
Unfortunately, this means we were unable to completely cover the full orbit
during the six nights of \ucam\ observations.

\begin{figure*}
\centering
\includegraphics[width=15cm]{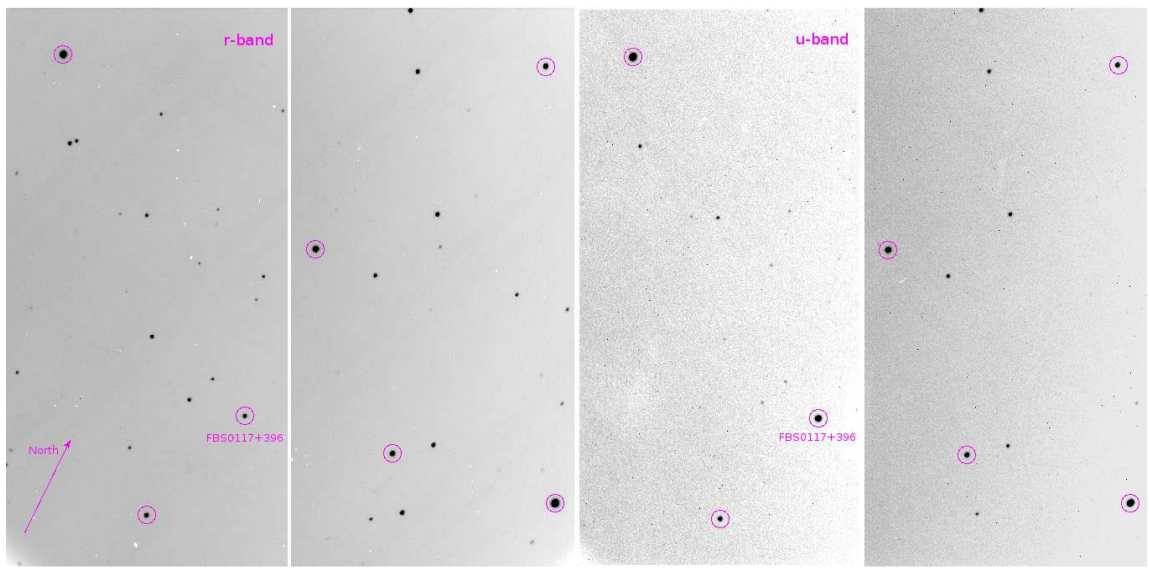}
\caption{
\ucam\ setup frame for the field of \target. The two images on the
left shows average frames from the left and right CCD outputs on the
red camera, and the right pair of frames from the u-band camera.
The target is labeled, and the
six reference stars used for the differential photometry are encircled.
}
\label{fig:targets}
\end{figure*}

The photometric data were processed by standard overscan bias-level subtraction
and flat-fielding, and the light curves extracted by aperture photometry.
For the \ucam\ data, we used the provided pipeline \citep{ultracam}, and
for the rest of the data we used our own RTP program \citep{ostensen01a}.
The extracted photometry was corrected for extinction, using suitable
extinction coefficients for the different bands. From these time series,
differential photometry was computed by dividing the target data series with that
of the sum of six reference stars with brightness comparable to the target.
The target and the reference stars are identified in Fig.~\ref{fig:targets}.
The \ucam\ data were observed using exposure times of $\sim$3 seconds for the $r$ and
$g$ channels, and twice of that for the $u$ band. The shorter exposure time for the
visual bands were required to avoid saturating the red reference stars.
For the white-light data shown in Fig.\,\ref{fig:lc_all} the \ucam\ frames
for all three bands were coadded to the sampling of the $u$-band. 

In order to best extract the $g$-mode pulsations we attempted to orbit
correct the differential light curves by fitting a function
of the form
\begin{eqnarray}
f(t) & = & A \cos \left( 2\pi \cdot {{t-T_0}\over{P}} \right) +
            B \cos \left( 4\pi \cdot {{t-T_0}\over{P}} \right) + C
\label{eq:lc}
\end{eqnarray}
where $P$ is the period, $t$ is the timestamp of each observation,
$T_0$ is first time of maximum light (zero phase) in the dataset,
and $A$, $B$ and $C$ are the fitting parameters. 
The form of this function approximates well the irradiation effect in
the similar \kep\ binaries \citep{kawaler10b}, where one only sees the orbital
frequency and its first harmonic even in near-continuous datasets spanning
a year or more. Using the cosine form with no phase offset ensures that the
maximum of the harmonic coincides exactly with the maximum of the orbital
period, giving sharp maxima and broad minima, as observed.
Thus, when using the period and phase from the radial-velocity solution,
with phase corresponding to that of a sine fit (as in Fig.~\ref{fig:rvplot}),
zero phase corresponds to the point in the orbit where the primary is at
its closest point relative to the observer, which is where we see maximum
light for the reflection effect.


\begin{table}
\begin{center}
\caption{Detected frequencies, amplitudes and phases.}
\label{tbl:freqs}\small
\begin{tabular}{lrrrrrrl}
\hline\hline\noalign{\smallskip}
   \multicolumn{1}{c}{ID} &
   \multicolumn{1}{c}{Freq.} &
   \multicolumn{3}{c}{Amplitude} &
   \multicolumn{1}{c}{Phase} \\
   &
   \multicolumn{1}{c}{[\uHz]} &
   \multicolumn{3}{c}{[mma]} &
   \multicolumn{1}{c}{[cyc]} \\
\noalign{\smallskip} \hline \noalign{\smallskip}
& & \multicolumn{1}{c}{$r$} & \multicolumn{1}{c}{$g$} & \multicolumn{1}{c}{$u$} & \\
\noalign{\smallskip} \hline \noalign{\smallskip}
$f_{1}$ & 2965.5(2) &  7.91(10) & 8.70(10) &12.45(15) & 0.045(2) \\
$f_{2}$ & 2636.4(2) &  2.85(10) & 3.05(10) & 4.32(15) & 0.324(5) \\
$f_{A}$ &  115.5(2) &  1.67(10) & 1.35(10) & 2.07(15) & ---\,$^a$ \\ 
$f_{B}$ &  171.2(2) &  0.80(10) & 0.75(10) & 1.46(15) & 0.79(3) \\ 
$f_{C}$ &  208.5(2) &  1.48(10) & 1.43(10) & 1.91(15) & 0.67(1) \\ 
$f_{D}$ &  260.1(2) &  0.55(10) & 0.42(10) & 0.91(15) & 0.80(5) \\ 
$f_{E}$ &  264.1(2) &  0.82(10) & 1.36(10) & 0.99(15) & 0.50(3) \\ 
$f_{F}$ &  339.6(2) &  1.16(10) & 1.00(10) & 1.37(16) & 0.63(2) \\ 
$f_{G}$ &  352.9(2) &  0.83(10) & 0.94(10) & 1.02(16) & 0.62(3) \\ 
$f_{H}$ &  359.8(2) &  0.57(10) & 0.83(10) & 0.76(15) & 0.08(3) \\ 
\noalign{\smallskip} \hline
\end{tabular}
\end{center}
Notes: $^a$ -- Inconsistent phase in the three bands.
\end{table}

\begin{figure*}
\centering
\includegraphics[width=15cm]{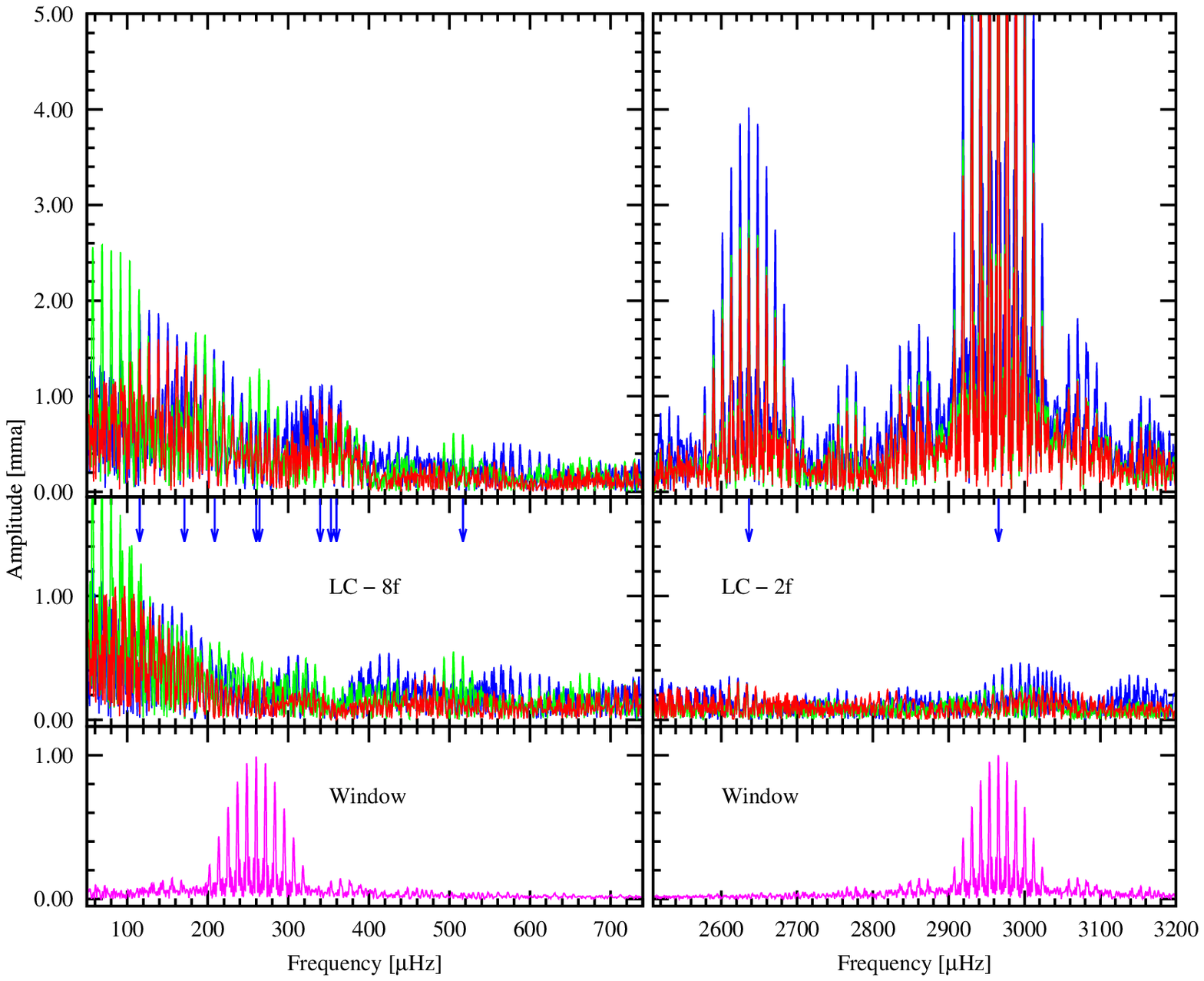}
\caption{
The amplitude spectrum of the orbit-subtracted \ucam\ light curve (LC), for the 
two frequency regions where pulsations are observed.
Upper panel: The amplitude spectra of the three
photometric channels (red: $r$-band, green: $g$-band, blue: $u$-band).
Middle panel: The same spectra after prewhitening the 10 frequencies 
listed in Table\,\ref{tbl:freqs} from the LC.
Lower panel: Window function for the two frequency regions.
}
\label{fig:ucam_ft}
\end{figure*}

Analysis of the white-light data produced consistent frequencies for the
two short-period frequencies, but the extended dataset proved to be too noisy
at low frequencies to examine the $g$-mode pulsation spectrum. We proceeded
to analyse the better quality \ucam\ data independently for each photometric band.

The detected frequencies, phases and the amplitudes for the three different bands
are listed in Table\,\ref{tbl:freqs}. To derive these we used the classic
non-linear-least-squares fitting method described in \citet{vuckovic06}. The
two short-period modes are trivial to fit, but the messy region below
500\,\uHz\ requires some iterations to produce consistent fits due to the forest of one-day
alias peaks. We first fitted each band individually, and then proceeded to 
fix frequencies that showed up consistently in all three bands
to the average of the fitting results for the individual
bands, in order to produce a ten-frequency fit where the frequencies where fixed
and the amplitudes and phases were adjusted. The resulting parameters are listed
in Table\,\ref{tbl:freqs}. There is significant residual power at low frequencies,
particularly for the $g$-band that is most likely due to variable sky transparency
or other effects not intrinsic to the target, so we did not attempt to fit
any signal below 100\,\uHz. The shortest frequency that was consistently found
in all bands, $f_A$\,=\,115.5\,\uHz, has a different phase in the $g$-band
relative to the two other bands. We cannot say if this is due to an
intrinsic frequency being affected by the higher noise at low frequencies in
that band, or if that period is a spurious signal due to changing observing
conditions. That it has consistent phase
in $u$ and $r$ indicates that it is most likely not just noise, but 
atmospheric effects cannot be ruled out.
For pulsators at the hot end
of the V1093-Her instability strip, \kep\ observations show typical cutoffs
at $\sim$100\,\uHz\ \citep{reed10a}, so we do not expect any of the residuals
below that limit to be intrinsic to the star.
For the other seven long-period modes the fit
converged with consistent phases in all three bands.
However, the amplitudes do not always follow the trend one would normally expect
for pulsations, where the amplitudes should decrease with wavelength.
Instead we see that the $g$-band amplitudes are lower than the $r$-band ones for
five of eight of the modes, and the $u$-band amplitude is not the highest for
two of the eight, but in all cases these trends are on the 1-$\sigma$ level.
This may be just a noise artifact, possibly combined
with atmospheric effects and dilution from the irradiated secondary,
which contaminates the signal more at longer wavelengths.

For the short-period modes the expected pattern comes out clearly above the
noise.
The amplitude ratios of the $f_1$ and $f_2$ are
($r$/$u$,$g$/$u$)\,=\,(0.64,0.70) and (0.66,0.71), which both have slightly 
higher contrast than ratios computed for a set of evolutionary models with
comparable atmospheric parameters from \citet{bloemen13}, but well
within the errors on the amplitude measurements.
Since $f_1$ and $f_2$ are too close together to have the same harmonic
degree, $\ell$, the simplest assumption is that they are $\ell$\,=\,0 and
$\ell$\,=\,1.
If they had the same intrinsic power, visibility would always
make the $\ell$\,=\,0 the highest amplitude mode, but 
the slightly higher amplitude ratio for $f_2$ than $f_1$
is not consistent with this picture.
Rather, the amplitude ratio is roughly of the order one would expect if
$f_1$ is the $\ell$\,=\,1 mode.

A Fourier transform of the multi-colour data, before and after prewhitening,
is shown in Fig.~\ref{fig:ucam_ft}.

\begin{figure}[t]
\centering
\includegraphics[width=\hsize]{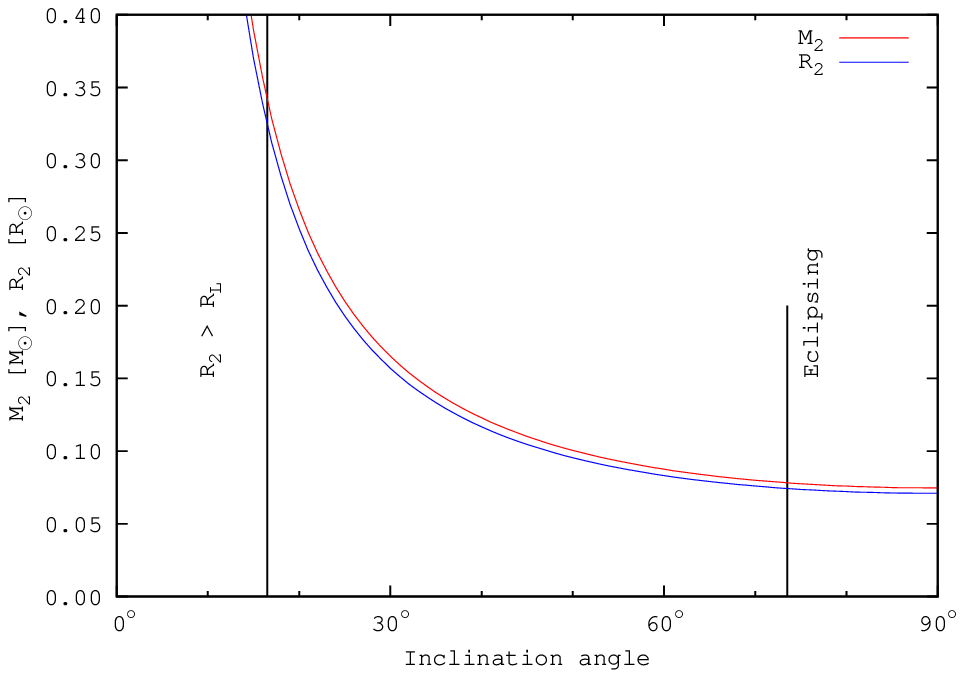}
\includegraphics[width=\hsize]{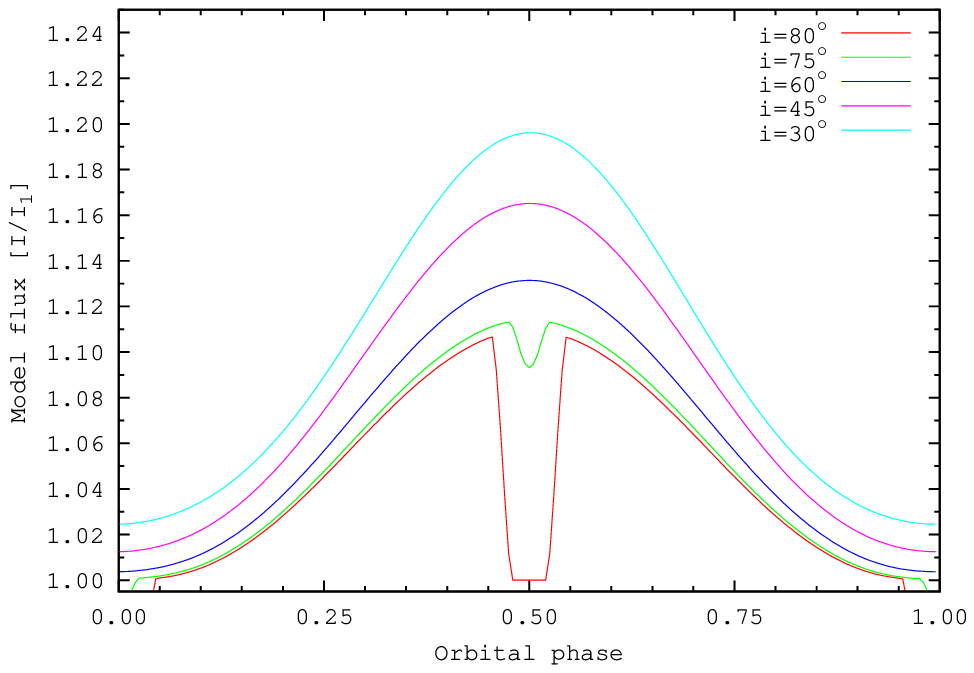}
\includegraphics[width=\hsize]{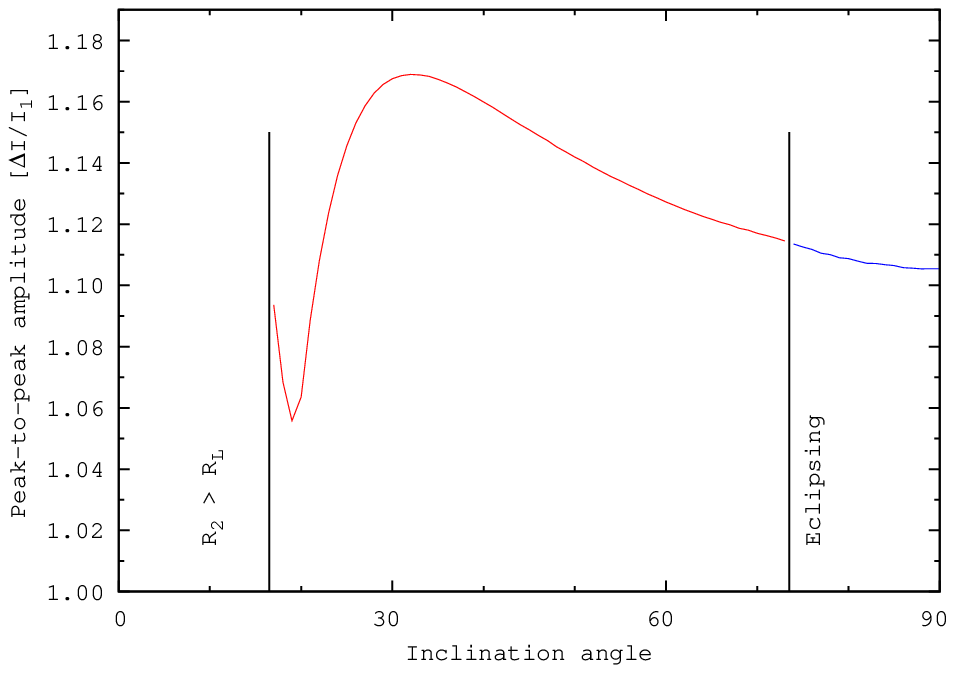}
\caption{
Upper panel: The mass and radius of the secondary component as a function
of inclination angle, as inferred by the mass function and the M/R relationship
for M-dwarfs.
Middle panel: Model light-curves for a few inclination angles, computed using
the radius relationship from the upper panel.
$I_1$ is the model flux of the
primary alone, and $I$ is the total model flux from the system.
Lower panel: Peak-to-peak amplitude of the reflection effect for all inclination
angles.
$\Delta I$ is the difference in model flux at phase 0 and phase 0.5,
when ignoring eclipses.
}
\label{fig:model_lc}
\end{figure}

\section{Modelling}

In the absence of eclipses it is not possible to determine the orbital inclination
from the light curves of sdB+dM binaries.
In order to investigate the possible ranges that the inclination angle could 
realistically take, we performed some simple light-curve modelling.
As starting assumptions we take the orbital period
and the mass function 
determined in Section\,\ref{sect:rv}. We assume the mass of the sdB to be
that of a canonical EHB star ($M_1$\,=\,0.48\,\msol) and use the mass-radius (M-R)
relationship for M-dwarfs from \citet{baraffe98}.
For each inclination angle, we get from the mass function a value for $M_2$, and from
the M-R relationship, we get the radius of the secondary. When we plot $M_2$ and
the corresponding radius, $R_2$ as a function of inclination angle, we get the
curves in the upper panel of Fig.\,\ref{fig:model_lc}. Inclinations lower than 16$^\circ$
are excluded as the secondary would fill its Roche lobe.

We can make some simple light curve models just by assuming
that each point on the secondary reflects light proportionally to the
geometrical fraction of the sky at that point that is covered by the primary
(including distance and horizon effects).
We only use spherical geometry, taking into account that the primary gets 
smaller when seen from points that are further away from the substellar point,
and bisected by the horizon as one approaches the edge of the irradiated
hemisphere.
We also take into account limb-darkening according to the square-root law with
coefficients from \citet{claret11}, as well as eclipses, to produce
light curves as shown in the middle panel of Fig.\,\ref{fig:model_lc}.
In reality the atmosphere is heated by the UV irradiation from the hot secondary,
and in order to estimate colour effects we would have to make assumptions
regarding the temperature distribution of the irradiated atmosphere. The problem
of degeneracies in such solutions still hamper adequate modelling even in the
presence of eclipses, so we will not attempt to produce a complete solution here.
Thus, the model light-curves shown here include geometric effects only, and
are therefore monochromatic. This is good enough for the eclipse effects, but
note that the peak amplitude of the irradiation effect is arbitrary, although
we can still make conclusions about how the irradiation effect varies with
inclination angle.
But it is interesting to note that the amplitude of the irradiation effect is
predicted to rise rapidly with decreasing inclination angle, as can be seen
in the middle and lower panels of Fig.\,\ref{fig:model_lc}. The amplitude
peaks around 22$^\circ$, due to limb-darkening effects. In models without
limb darkening for the secondary, the amplitude does not peak before the
Roche limit is reached. Note also that we chose limb-darkening coefficients
for the hot side of the secondary to be the same as for the sdB, which
is a completely arbitrary choice. In fact, since the atmosphere is inverted
(its temperature is highest in the outermost layers), its spectral lines may be
in emission as demonstrated for the companion of AA\,Dor by \citet{vuckovic08},
one might expect limb brightening rather than limb darkening, which would give
a more extreme inclination/amplitude relationship.  To our knowledge, detailed
limb-darkening (-brightening) parameters for irradiated M-dwarf atmospheres
have not been computed.

The main result here is that the amplitude of the irradiation effect increases
with decreasing inclination angle for most of the possible parameter space.
This is rather counter-intuitive, but is caused by the constraint on
$K_1$ from spectroscopy and the relationship between $i$ and
and $M_2$ and $R_2$ that follows. Unfortunately, the absolute value
of the amplitude of the reflection effect has too many associated uncertainties
to be used to constrain the inclination angle from this effect.
We hope that with some modelling of the limb-darkening effects associated
with irradiated atmospheres, this situation can be remedied.

\begin{figure}[t]
\centering
\includegraphics[width=\hsize]{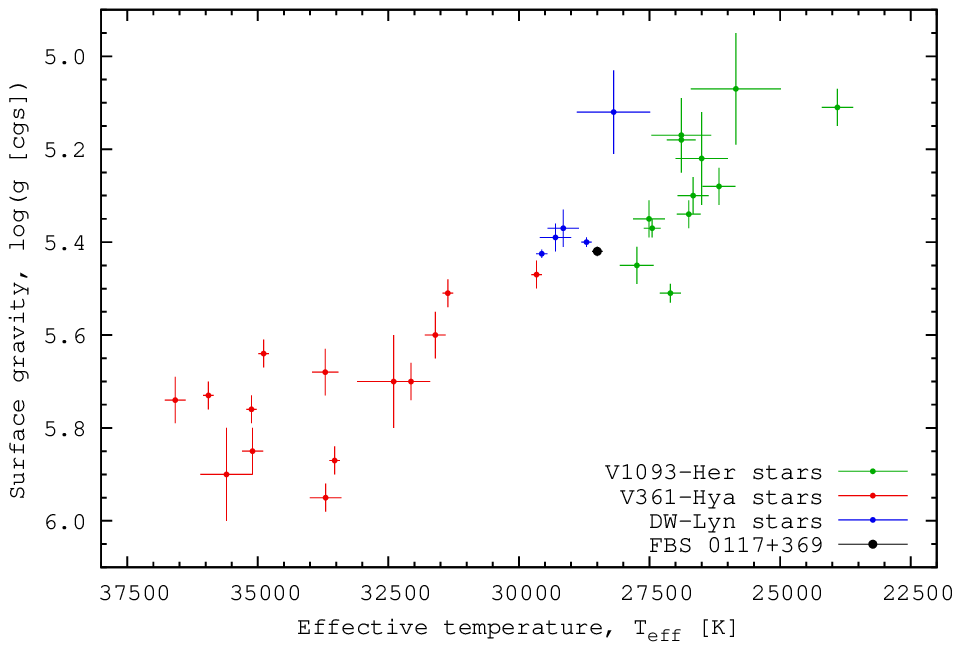}
\caption{
\teff/\logg\ diagram. \target\ falls neatly with the close group of
hybrid (DW-Lyn type) sdBV stars. The V1093-Her stars are the pulsators in
the \kep\ sample with parameters from \citet{ostensen10b,ostensen11b},
and the other stars are summarised in \citet{sdbnot}, but we have
only included those that have been fitted on the same LTE grid as
described in Section\,\ref{sect:spfit}.
}
\label{fig:tgplot}
\end{figure}

\section{Discussion and conclusions}

Subdwarf-B stars must experience extreme mass loss in order to expel
almost their entire hydrogen envelopes on the red-giant branch.
Surveys have revealed fractions of short-period binaries
that range from 69\% \citep{maxted01} to 39\%
\citep{napiwotzki04a}, depending on how the sample was selected.
A large fraction of the remaining stars are binaries
in long-period systems, where the mass ratio was inverted during
mass transfer, allowing the orbit to expand to exceedingly long
periods, but some definitely single stars are known to exist
providing an enduring mystery as to their formation \citep{ostensen09}.
Among the short-period binaries, sdB stars with M-dwarf companions are
rare compared to systems with unseen companions (presumably white
dwarfs); about 1/5 in the compilation of \citet{geier11a}.
Short-period pulsations are only found in $\sim$10\% of sdB stars that
are placed at the hot tip of the EHB \citep{sdbnot}, and from that
sample of 49 stars plus the recent discoveries,
KIC\,10139564 \citep{kawaler10a}, J08069+1527 \citep{baran11a},
J20136+0928 \citep{ostensen11a}, J06398+5156 \citep{vuckovic12},
and the current \target\ bringing the total up to 54,
we expect about half (27) to be in short-period binaries, and of
those 1/5 should be sdBV+dM systems. Thus, finding a fourth system
still leaves us one short of the predicted five. Although a discrepancy
of one is not statistically significant, when considering that
many of the 54 short-period sdBVs have still not
been carefully checked for RV variations, it is not excluded that
more may remain to be discovered.

\target\ falls nicely in the clump of hybrid pulsators in the
\teff/\logg\ diagram (Fig.\,\ref{fig:tgplot}).
Up to now it appeared that all the DW-Lyn pulsators that
cluster together in the \teff/\logg\ diagram were single stars.
2M1938+4603 is an exception, of course, as it is included in the
clump of blue points here, but since its pulsation spectrum is
quite different from the DW-Lyn stars (it has no high-amplitude
$p$-modes) it should perhaps not be termed as such. The V361-Hya
star that mingles in on the hot side of the clump is
KL\,UMa (=\,Feige\,48) which is a binary with an unseen companion
with a rather long orbital period ($P$\,=\,9\,h) \citep{otoole04}.
However, hybrid pulsations have not been detected in this pulsator.

\target\ is particularly interesting as its orbital period at $\sim$6\,h
is longer than that of the stars that are known to be tidally locked
(NY\,Vir; $P$\,=\,2.4\,h, 2M1938+4603; $P$\,=\,3\,h) and the \kep\
binaries of \citet{pablo11} which at $P$\,$\approx$\,10\,h are not.
But whether or not \target\ is in a tidally locked orbit cannot be determined
from the current data.
The observed velocity required for a locked state depends on the inclination
angle, and for $i$\,=\,45$^\circ$ and $R_1$\,=\,0.22\,\rsol, we would expect
$v\sin i$\,=\,44\,km/s, which is not detectable in our low S/N individual spectra
that has a resolution of $\sim$75\,km/s at best.
It might have been detectable in some of our mean spectra if  we could have
added up the individual observations while correcting for the orbital Doppler shifts, thereby
beating down the noise. However, the problems with the wavelength calibration discussed in
Section~\ref{sect:rv} hobbles this approach.
Furthermore, the pulsation spectrum is too simple to give any constraints
on the rotation from rotationally split multiplets. Thus,
we are not able to constrain the value of the rotation from the available data.
A high-resolution spectrum of \target\ would help with resolving this issue.

\begin{acknowledgements}

Based on observations collected at the Nordic Optical Telescope ({\scriptsize NOT}), 
the William Herschel Telescope (WHT) and the Mercator Telescope,
all on the island of La Palma, Spain, and at the
Centro Astr\'onomico Hispano Alem\'an ({\scriptsize CAHA}) at Calar Alto, Spain, and
at the Observatorio Astron\'omico Nacional (OAN) at San Pedro M\'artir, Mexico.
We gratefully acknowledge using spectroscopic data from the database of the
Sloan Digital Sky Survey (SDSS).

The research leading to these results has received funding from the European
Research Council under the European Community's Seventh Framework Programme
(FP7/2007--2013)/ERC grant agreement N$^{\underline{\mathrm o}}$\,227224
({\sc prosperity}),
from the Research Council of KU~Leuven grant agreement GOA/2008/04,
and from the Fund for Scientific Research
of Flanders (FWO), grant agreements G.0332.06 and G.0470.07, and G.0C31.13. 
The CCD of the Merope\,{\sc ii} camera is part of a set developed by E2V in the
framework of the Eddington space mission project and are owned by the
European Space Agency; they were
offered on permanent loan to the Institute of Astronomy of KU\,Leuven, Belgium,
with the aim to build and exploit an instrument for asteroseismology research to
be installed at the Mercator telescope.
Conny Aerts is grateful to Giuseppe Sarri and Fabio Favata for their
support and help in the practical implementation of the ESA loan agreement with
KU\,Leuven.

SG, AT, EZ and travel to Calar Alto were supported by the
Deutsche Forschungsgemeinschaft (DFG) through grants HE1356/49-1, HE1356/45-1
HE1356/45-2, HE1356/53-1, and HE1356/64-1.

VS acknowledges funding by the Deutsches Zentrum f\"ur Luft- und Raumfahrt 
(grant 50 OR 1110) and by the Erika-Giehrl-Stiftung.
LFM acknowledges financial support from the UNAM under grant PAPIIT 104612.

SB is supported by the Foundation for Fundamental Research on Matter (FOM),
which is part of the Netherlands Organisation for Scientific Research (NWO).

\end{acknowledgements}
\bibliographystyle{aa}
\bibliography{sdbrefs}

\end{document}